\documentclass[pss]{wiley2sp}
\usepackage{units}
\usepackage{hyperref}
\usepackage[T1]{fontenc}
\usepackage[latin1]{inputenc}
\usepackage{amssymb}
\usepackage{amsmath}
\usepackage{enumitem}
\usepackage{amsfonts}
\usepackage{color}
\usepackage{graphicx}
\usepackage[numbers]{natbib}
\usepackage{bm}
\begin{document}

\title{Using the Callaway model to deduce relevant phonon scattering processes: The importance of phonon dispersion}
\author{%
  Matthias Schrade\textsuperscript{\Ast,\textsf{\bfseries 1}},
  Terje G. Finstad\textsuperscript{\textsf{\bfseries 1}},
  }

%E-mail-address of corresponding author
\mail{e-mail
  \textsf{matthias.schrade@smn.uio.no}}

%\author{Matthias Schrade}
%\email{matthias.schrade@smn.uio.no}
%\affiliation{Department of Physics, Centre for Materials Science and Nanotechnology, University of Oslo, Sem S{\ae}landsvei 26, 0371 Oslo, Norway}
%\author{Terje G. Finstad}
%\affiliation{Department of Physics, Centre for Materials Science and Nanotechnology, University of Oslo, Sem S{\ae}landsvei 26, 0371 Oslo, Norway}
\institute{%
  \textsuperscript{1}\,Department of Physics, Centre for Materials Science and Nanotechnology, University of Oslo, Sem S{\ae}landsvei 26, 0371 Oslo, Norway}
\date{15. July 2018}
\abstract{\bf
The thermal conductivity $\kappa$ of a material is an important parameter in many different applications. Optimization strategies of $\kappa$ often require insight into the dominant phonon scattering processes of the material under study. The Callaway model is widely used as an experimentalist's tool to analyze the lattice part of the thermal conductivity, $\kappa_l$. {\color{black}Here, we investigate how deviations from the implicitly assumed linear phonon dispersion relation affect $\kappa_l$ and in turn conclusions regarding the relevant phonon scattering processes.} As an example, we show for the half-Heusler system (Hf,Zr,Ti)NiSn, that relying on the Callaway model in its simplest form has earlier resulted in a misinterpretation of experimental values by assigning the low measured $\kappa_l$ with unphysically strong phonon scattering in these materials. Instead, we propose an implementation of more realistic phonon dispersion curves, combined with empirical expressions for typical phonon scattering processes, which leads to far better quantitative agreement with both theoretical and experimental values. {\color{black}This method can easily be extended to other materials with known phonon dispersion relations.} 
}

\maketitle
\section{Introduction}
The thermal conductivity $\kappa$ is an important material property in many applications operating at room temperature or above, including for example heat management in integrated circuit design \cite{Tuckerman1981}, protective coatings \cite{Cao2004}, and thermoelectrics \cite{Snyder2008}. $\kappa$ usually has contributions from electronic charge carriers, $\kappa_e$, and lattice vibrations, $\kappa_l$. For a given material, $\kappa_l$ can be modified by microstructural engineering and the control of phonon scattering centers, while $\kappa_e$ is rigidly connected to the electrical conductivity via the Wiedemann-Franz law. For applications, where both the thermal and electrical conductivity are important for the optimization of performance, only $\kappa_l$ can be independently optimized, and thus can become of prime importance for optimizing the thermal conductivity \cite{Li2010,Vineis2010}.
{\color{black}Deducing the relevant microscopic phonon scattering processes from the experimentally measured, macroscopic $\kappa_l$ is often challenging, but required in order to develop efficient strategies to optimize $\kappa_l$ for the given application.}

Density functional theory (DFT) based {\itshape ab initio} methods scale unfavorably with the system size, due to the need for higher order perturbation terms, and therefore require high, cost-intensive computational power. Despite enormous progress during the last years {\color{black}\cite{Jeong2011,Bjerg2014,Wang2017,Tian2012,Togo2015,Li2014,Plata2017,Carbogno2017,Nath2017,Toher2017}}, these methods are still mainly used to describe specific, well defined material systems, so that fast, routine analysis of experimental data is not feasible at the moment. Therefore, experimental data is still often analyzed using simplified models developed decades ago.

A much used simplified model is the "Callaway model", which is based on the Debye treatment of phonons and calculates the lattice thermal conductivity of a material via an effective phonon relaxation time $\tau$ \cite{Callaway1959}.\footnote{Here, we study $\kappa_l$ at temperatures equal or larger than the Debye temperature, where $U-$ are considered to dominate over $N-$processes. We therefore neglect $N-$processes in the present paper.}
In its common form, the Callaway model is implicitly assuming a linear $\omega(k)$ phonon dispersion throughout the 1$^\text{st}$ Brillouin zone (1.BZ). Under this assumption, the group velocity, $v_g=\text{d}\omega/\text{d}k$, is constant and identical to the phase velocity, $v_p=\omega/k$, and sound velocity, $v_s=\text{d}\omega/\text{d}k\,_{k\rightarrow 0}$.

However, real materials all show deviations from the simple idealized dispersion, and the goal of the present paper is to demonstrate the errors introduced in the calculated lattice thermal conductivity by using the standard Callaway model. By implementing a simple phonon dispersion into the model, we show that it gives a significantly better agreement with results obtained by experimental and higher-level theoretical methods. {\color{black}This generalization can easily be extended to other material systems, to deduce the relevant phonon scattering processes from experimental data, without the computational cost of an {\itshape ab initio} based theoretical analysis.}
\begin{figure}[tb]
\centering
\includegraphics[width=0.48\textwidth]{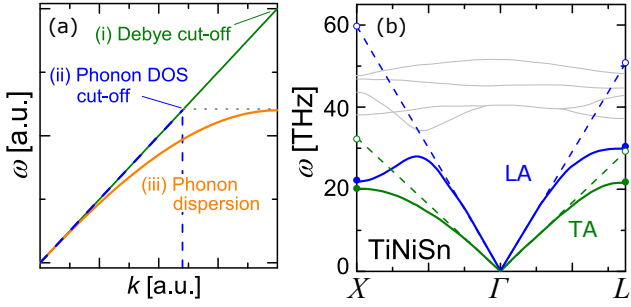}
\caption{(a) Schematic of the three different approximations to the Callaway model used here, as explained in the text. (b) Phonon dispersion for TiNiSn along selected high symmetry directions as calculated by DFT. Optical modes are greyed out for clarity. The linear extrapolation around $\Gamma$ leads to a significant overestimation of cut-off frequencies for both longitudinal (LA) and transversal acoustic (TA) phonon branches.}
\label{fig_dispersion_and_sketch}
\end{figure}
\section{Calculation of thermal conductivity}\label{experimental_computational}
W{\color{black}ithin the limits of the relaxation time approximation and} neglecting the small influence of $N$-processes at elevated temperatures, $\kappa_{l,i}$ of the $i$-th acoustic phonon mode is generally given as {\color{black}\cite{Ward2010}}
\begin{equation}
\kappa_{l,i}=\frac{1}{3}\int_0^{\omega_\text{Max}}{C_{V,i}(\omega)\tau_i(\omega)v_{g,i}(\omega)^2\text{d}\omega}
\label{eq_1}
\end{equation}
where $C_{V,i}(\omega)$ is the spectral specific heat, $\tau_i(\omega)$ the relaxation time, and $v_{g,i}(\omega)$ the group velocity of this mode. All phonons up to a maximum cut-off frequency $\omega_\text{Max}$ are considered in the integral. The total lattice thermal conductivity is then obtained as a sum over the contributions from the two transversal and the one longitudinal acoustic modes.\cite{Morelli2002,Asen-Palmer1997} As many parameters entering the calculations are only reported in the literature as an average over the three acoustic modes, we here use these average values instead {\color{black} to ease comparison with reported results}. For example, the average sound velocity $v_s$ is related to the velocity of the longitudinal, $v_l$ and tranversal modes, $v_t$ via $v_s=\left[1/3\cdot(v_l^{-3}+2v_t^{-3})\right]^{1/3}$ \cite{Xie2013}. Adding the contributions of the three acoustic phonon branches, using the Debye expression for the specific heat, and $x=\hbar\omega/k_BT$, Eq. \eqref{eq_1} can be rewritten as \cite{Toberer2011,Morelli2002}
\begin{equation}
\kappa_{l}=\frac{k_B}{2\pi^2}\left(\frac{k_BT}{\hbar}\right)^3\int_0^{\hbar\omega_\text{Max}/k_BT}{\tau(x)\frac{v_{g}(x)x^4e^x}{v_p(x)^2(e^x-1)^2}\text{d}x}
\label{eq_2}
\end{equation}
When further assuming a linear dispersion, i.e. $v_g=v_p=v_s$, Eq. \eqref{eq_2} simplifies to the commonly used "Callaway expression" \cite{Tritt2004}
\begin{equation}
\kappa_{l}=\frac{k_B}{2\pi^2v_s}\left(\frac{k_BT}{\hbar}\right)^3\int_0^{\hbar\omega_\text{Max}/k_BT}{\tau(x)\frac{x^4 e^x}{(e^x-1)^2}dx}
\label{kappa_callaway}
\end{equation}
\footnote{Recently, Allen pointed out a correction of the original Callaway formula \cite{Allen2013}. Since this modification only addresses the $N-$process correction term of the total $\kappa_l$, it is here neglected for the discussion of thermal conductivity at elevated temperatures.}

As a model system, we choose here the half-Heusler $X$NiSn, ($X=\text{Hf, Zr, Ti}$), which has attracted considerable interest due to the combination of environmental abundance, non-toxicity, and promising thermoelectric properties \cite{Krez2016}. In order to decrease the high $\kappa_l$ of these alloys, mass disorder on the $X$ site has been widely investigated \cite{Sakurada2005,Culp2006,Geng2014,Liu2015,Xie2012b}. Xie {\itshape et al.} concluded on a dominating role of electron-phonon scattering in a wide phonon frequency range to describe the experimentally obtained $\kappa_l$ of these materials \cite{Xie2012b}. Their analysis was based on Eq. \ref{kappa_callaway} in combination with empirical expressions for the scattering times $\tau_i$ of the individual scattering processes. However, the scattering rate for electron-phonon scattering should scale with the concentration of free electrons in the system. If electron-phonon scattering was a dominant contribution to the total $\kappa_l$ of $X$NiSn, one should observe a significant reduction of $\kappa_l$ with increasing charge carrier concentration. However, in systematic doping studies, $\kappa_l$ shows no clear trend with varying carrier concentration \cite{Sakurada2005,Xie2012a,Yu2012}.
On the contrary, we have recently reported the good agreement of experimental values of $\kappa_l$ for $X$NiSn and {\itshape ab initio} DFT calculations, without taking into account electron-phonon scattering \cite{Schrade2017}. In the following, we will therefore use those reported DFT values \cite{Eliassen2017} as an experimentally confirmed reference for the analysis in the present paper.

In order to calculate the lattice thermal conductivity via Eq. \eqref{eq_2}, one needs (a) a cut-off frequency $\omega_\text{Max}$ as an integration limit, (b) an expression for the specific form of the phonon dispersion relation $\omega(k)$, and (c) an expression for the spectral phonon relaxation time $\tau(\omega)$.

An obvious shortcoming of the frequently used Eq. \eqref{kappa_callaway} is the assumed linear phonon dispersion throughout the 1. BZ. This was originally justified, as Callaway's model was intended to describe $\kappa_l$ at low temperatures, where most phonons populate states in the linear regime of the dispersion curves around the center of the 1.BZ, $\Gamma$ \cite{Callaway1959}. However, at elevated temperatures, also phonon modes outside the linear regime of the dispersion relation contribute to the thermal conduction. 
%Assuming a linear phonon dispersion throughout the Brillouin zone, the cut-off frequency $\omega_\text{Max}=\omega_\text{Debye}$ can be related to the sound velocity by $\omega_\text{Debye}=v_s(6\pi^2V)^{1/3}$, where $V$ is the volume per atom in the structure.
Estimating the integration cut-off by extrapolating the linear region of the phonon dispersion curves to the boundary of the 1.BZ will thus generally result in too high $\omega_\text{Max}$ and thus too high $\kappa_l$. Instead, it was suggested to use the phonon frequency at the zone boundary as a cut-off in Eq. \eqref{kappa_callaway}, which can be obtained by, for example, inelastic neutron diffraction experiments or computational studies \cite{Morelli2002,Zhang2016}.

As a second requirement to calculate $\kappa_l$ by Eq. \eqref{eq_2}, one has to assume a certain shape of the phonon dispersion relation, providing expressions for $v_g(x)$ and $v_p(x)$. Here, we calculate $\kappa_l$ by three different methods, schematically shown in Fig. \ref{fig_dispersion_and_sketch} (a):
\begin{enumerate}[label=(\roman*) ]
\item{ The "Debye cut-off" model, i.e. using linear phonon dispersion throughout the Brillouin zone, with $v_g=v_p=v_s=v$.}
\item{ The "Phonon DOS cut-off" model: As in (i), but limiting the integral to frequencies below the phonon frequency at the Brillouin zone boundary, $\omega_\text{Max}$, as obtained from computational phonon dispersion curves. Note that this approach still assumes the identity of sound, group, and phase velocity for all phonon modes considered in the integral.}
\item{ The "Phonon dispersion" model, which approximates the average phonon dispersion by a simple polynomial, $\omega(k)=a\cdot k^3+b\cdot k^2+v_s\cdot k$. The coefficients $a$ and $b$ for the different compositions $X$NiSn are obtained from the condition $\omega(k=k_\text{Max})=\omega_\text{Max}$ and $\text{d}\omega/\text{d}k\,_{k=k_\text{Max}}=0$. From the approximated $\omega(k)$, the frequency dependent $v_g(\omega)$ and $v_p(\omega)$ in Eq. \eqref{eq_2} can be easily obtained.}
\end{enumerate}

Lastly, in order to calculate the thermal conductivity for the three different models using Eq. \eqref{eq_2}, information on the frequency dependent relaxation time $\tau(\omega)$ is needed. According to Matthiessen's rule, the total relaxation time $\tau(\omega)$ is related to the relaxation time $\tau_i(\omega)$ of the $i$-th scattering process via $\tau(\omega)^{-1}=\sum_i{\tau_i(\omega)^{-1}}$.
In the present calculation, we consider three different phonon scattering mechanisms: Umklapp , mass disorder, and grain boundary scattering. Importantly, this is the same set of phonon relaxation processes contributing to the total relaxation time $\tau$ as used in earlier DFT calculations with good agreement with experimental values \cite{Eliassen2017,Schrade2017}.
For the relaxation times of the different processes, we adapt commonly used empirical expressions, summarized in Table \ref{scattering_processes}.
There, $\gamma$, $\Theta_\text{Max}=\hbar\omega_\text{Max}/k_B$, $d$, $\bar{m}$, $\bar{M}$, $c_i$, and $m_i$ are the Gr\"uneisen parameter, the frequency cut-off, the average grain size, the average atomic mass on the $X$-sublattice, the average mass of all atoms, the relative concentration of Hf, Zr, or Ti, and their atomic mass, respectively.
$\gamma$ is related to the anharmonicity of the chemical bonds of the structure. Reported values of $\gamma$ for $X$NiSn vary in the range from $0.8-2$ depending on the employed method \cite{Hermet2014,Xie2013,Wee2012}, but do not show a significant variation with $X$ using the same methodology \cite{Toher2014}. Here, we use $\gamma=2$ for all compositions, but our results are not very dependent on this particular choice. For point-defect scattering, we also neglect the contribution due to strain variations, as it has been shown to be a minor contribution to $\tau_{PD}$ for $X$NiSn \cite{Geng2014}. 

We note that, in contrast to many previous studies \cite{Tiwari1971,Holland1963}, where the scattering strength of the individual mechanisms is obtained via fitting of experimental data, $\tau(\omega)$ is here obtained from independent, tabulated material properties.
\begin{figure}[tb]
\centering
\includegraphics[width=0.48\textwidth]{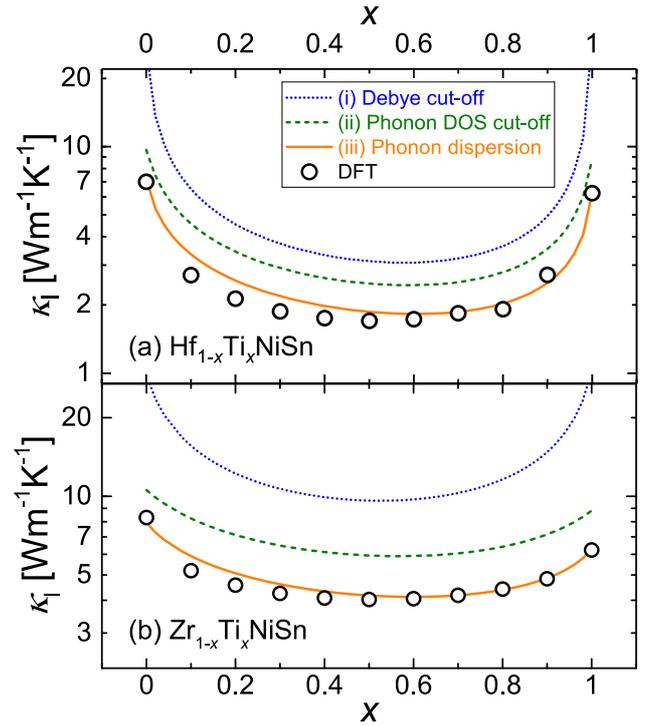}
\caption{The lattice thermal conductivity of Hf$_{1-x}$Ti$_x$NiSn (a) and Zr$_{1-x}$Ti$_x$NiSn (b), calculated by the three modifications of the Callaway model and compared to DFT-based results. Calculations are done at \unit[400]{K} and using an average grain size of \unit[100]{nm}. While all methods predict a similar qualitative behavior, with a strong reduction of $\kappa_l$ with increasing mass disorder, only the simple phonon dispersion relation (orange) reproduces the DFT-results in a quantitative way.}
\label{fig_kappa_l}
\end{figure}
\begin{figure}[tb]
\centering
\includegraphics[width=0.48\textwidth]{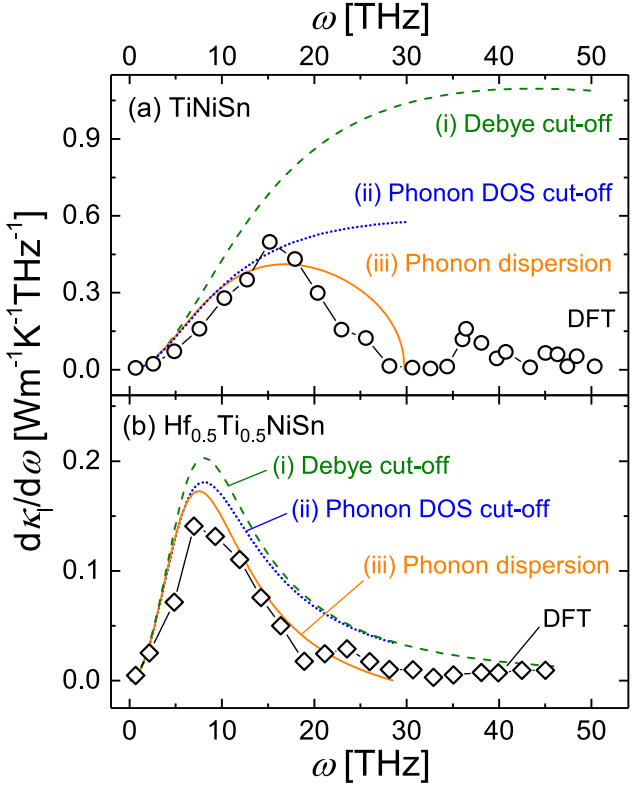}
\caption{(a) Spectral thermal conductivity for TiNiSn at \unit[400]{K} and for an average grain size of \unit[100]{nm} using the three methods described in the text and compared to earlier DFT-based results \cite{Eliassen2017}. The high cut-off frequency of model (i) extends far into the frequency range of optical phonon modes. Model (ii) leads to a significantly lower $\kappa_l$, but only model (iii) qualitatively reproduces the DFT result. (b) As in (a), but for Hf$_{0.5}$Ti$_{0.5}$NiSn. The difference between the three methods is less dramatic, but implementing a more realistic phonon dispersion into the Callaway model still leads to better agreement with the DFT-based reference.}
\label{fig_spectral_figure}
\end{figure}
\begin{table*}[tb]
\caption{Expressions for the relaxation rates of the different scattering processes considered here. Symbol explanation and further information is given in the text.}
\begin{tabular}{ccc}
\hline
Scattering Process&Scattering rate&Reference\vspace{0.0025\textwidth}\\
\hline
Umklapp&$\tau_U^{-1}(\omega)=\frac{\hbar\gamma^2T}{\bar{M}v_p(\omega)^2\Theta_\text{Max}}\exp\left(-\frac{\Theta_\text{Max}}{3T}\right)\times\omega^2$&Ref. \cite{Slack1964}\vspace{0.01\textwidth}\\
Point defect&$\tau_{PD}^{-1}(\omega)=\frac{V}{12\pi v_g(\omega)\cdot v_p(\omega)^2}\left(\left(\frac{\bar{m}}{\bar{M}}\right)^2\sum_{i=\text{Hf, Zr, Ti}}{c_i\left(\frac{m_i-\bar{m}}{\bar{m}}\right)^2}\right)\times\omega^4$&Refs. \cite{Yang2004,Klemens1955,Holland1963}\vspace{0.01\textwidth}\\
Grain boundary&$\tau_{GB}^{-1}(\omega)=v_g(\omega)/d$&Ref. \cite{Callaway1960}\\
\hline
\end{tabular}
 \label{scattering_processes}
\end{table*}

\begin{table}[tb]
\caption{The longitudinal $v_l$, transversal $v_t$, average sound velocity $v_s$, Debye cut-off $\omega_\text{Debye}$ and phonon DOS cut-off $\omega_\text{Max}$ for the three unmixed compounds. Values for the mixed compositions are obtained by linear interpolation of the unmixed values. There is a good agreement with  experimental values obtained for samples of similar composition (\unit[$\pm5$]{\%}).\cite{Xie2013}}
 \begin{tabular}{cccccc}
\hline
$X$ & $v_l$& $v_t$& $v_s$& $\omega_\text{Debye}$ & $\omega_\text{Max}$\\
&\multicolumn{3}{c}{[ms$^{-1}$]}&\multicolumn{2}{c}{[THz]}\vspace{0.002\textwidth}\\
\hline
Ti&2991&5513&3337&50.3&30.0\\
Zr&2878&5437&3217&47.0&30.6\\
Hf&2575&4793&2875&42.2&27.4\\
\hline
\end{tabular}
 \label{tab_sound_velocity}
\end{table}

\section{Results and discussion}
From the available DFT-based phonon dispersion curves, Fig. \ref{fig_dispersion_and_sketch} (b), for the three unmixed compounds, ($X=\text{Hf, Zr, Ti}$) \cite{Page2015,Eliassen2017}, we obtained the sound velocity $v_s$ from a linear fit around the $\Gamma$-point, as shown in Table \ref{tab_sound_velocity}. From the calculated dispersion curves, we can then quantify the overestimation of the cut-off frequency in Eq. \eqref{eq_2}, when erroneously using the linear phonon dispersion throughout the 1.BZ, instead of the real phonon frequency at the zone boundary. Extrapolating the linear region of all phonon branches of \mbox{TiNiSn} leads to much higher phonon frequencies at the Brillouin zone boundary, $\omega_\text{Debye}$, than the frequencies obtained by DFT. This is most pronounced for the LA branch, but also extrapolating the TA branches results in an overestimation of the maximum frequency of around \unit[40]{\%}. Averaging over the three acoustic phonon branches and the different $k-$space directions, the average frequency at the zone boundary, $\omega_\text{Max}\approx 0.6\,\omega_\text{Debye}$, fairly independent of the composition $X$, {\itshape cf.} Table \ref{tab_sound_velocity}.

Fig. \ref{fig_kappa_l} compares $\kappa_l$ calculated from the different models for different compositions $X$. As expected, in all models $\kappa_l$ is strongly reduced when introducing mass disorder on the $X$ site. However, both the "Debye cut-off" (i) and "Phonon DOS cut-off" (ii) model predict significantly higher values than the DFT reference data set, in particular for the unmixed compositions. In contrast, values calculated using the "Phonon dispersion" model (iii) are very similar to the DFT results for all compositions.
To understand these findings on a microscopic level, it is instructive to examine the spectral thermal conductivity $\text{d}\kappa_l(\omega)/\text{d}\omega$, instead of the integrated value \cite{Zhu2013}. While the spectral thermal conductivity is easily obtained from computational studies, experimental access is very challenging \cite{Minnich2011,Hu2015}, making it difficult to directly compare the modeled predictions with experimental reality. However, it is intuitively clear that phonons with $v_g(\omega)=0$ do not contribute to $\kappa_l$, in particular that $\text{d}\kappa_l/\text{d}\omega\,_{\omega\rightarrow\omega_\text{Max}}=0$. The spectral thermal conductivity for $X=\text{Ti}$ and $X=\text{Hf}_{0.5}\text{Ti}_{0.5}$ at \unit[400]{K} and for an average grain size of \unit[100]{nm} is shown in Fig. \ref{fig_spectral_figure}. For the DFT results, the dominant contribution to $\text{d}\kappa_l(\omega)/\text{d}\omega$ is -- as expected -- for phonon frequencies of the acoustic branches with $\omega\le\unit[30]{THz}$. In model (i), the cut-off frequency is significantly overestimated and, in the absence of point-defect scattering, high frequency phonons contribute significantly to the calculated $\kappa_l$, Fig. \ref{fig_spectral_figure} (a). Reducing the cut-off frequency improves the agreement with the DFT result (ii), but for both (i) and (ii) $\text{d}\kappa_l/\text{d}\omega\,_{\omega\rightarrow\omega_\text{Max}}\ne0$, as a consequence of the constant group velocity assumed in these models. Best qualitative and quantitative agreement is achieved for the simple phonon dispersion model (iii). For $X=\text{Hf}_{0.5}\text{Ti}_{0.5}$, Fig. \ref{fig_spectral_figure} (b), the large mass-disorder on the $X$ site brings down the contribution of high frequency phonons for all models, but implementing model (iii) still leads to a significant better agreement with the DFT reference data.
{\color{black}All three variations of the Callaway model investigated here ignore the contribution of optical phonon modes to $\kappa_l$. This is usually acceptable for most semiconductors, due to the relatively simple crystal structure and a small $v_g$ of the optical modes compared to the acoustic branches \cite{Tritt2004}. However, for materials with many atoms in the primitive unit cell, the high number of optical modes can compensate for the low $v_g$, so that these modes may carry a significant part of the heat flux in the material. In the case of those materials, we emphasize that one {\itshape a priori} cannot simply express the contribution of the optical modes like in Eq. \eqref{eq_1} and add them to the acoustic $\kappa_l$. The empirical scattering rates $\tau_i^{-1}$, as summarized in Table \ref{scattering_processes}, have been developed only taking into account acoustic modes, so that some influence of the optical branches, like scattering from or into these modes, might effectively already be incorporated in the expression of $\tau_i^{-1}$. For $X$NiSn, the contribution of optical modes to the total $\kappa_l$ as obtained from the full dispersion DFT calculation is below \unit[10]{\%}, Fig. \ref{fig_spectral_figure}, i.e. below typical accuracies of a Callaway-type analysis.}

Our results emphasize that the linear dispersion relation as assumed in Eq. \eqref{kappa_callaway} predicts a $\kappa_l$, which is much larger than values typically measured experimentally. To make values of $\kappa_{l}$ calculated by the Debye-Callaway model similar to experimental ones, different approaches have been reported in the literature. These include introducing additional phonon relaxation processes, such as electron-phonon scattering, reducing the effective relaxation time for low and medium frequency phonons \cite{Xie2013,Liu2015}, or using an empirical expression for $\tau^{-1}=A\omega^4+B\omega^2+v/d$, with $A$, $B$, and $d$ being adjustable parameters \cite{Geng2014,Petersen2015}. In the latter approach, determining $A$, $B$, and $d$ may well reproduce the experimentally measured $\kappa_l$, but the physical meaning of these parameters can be unclear. More severely, adding additional phonon scattering processes to the analysis changes the physical interpretation of the obtained results.

In addition, both these modifications only target the integrated $\kappa_l$, while the spectral $\text{d}\kappa_l(\omega)/\text{d}\omega$ would show finite values up to the Debye cut-off $\omega_\text{Debye}$, in disagreement with physical intuition and DFT results. As we show here, much better agreement for both $\kappa_l$ (Fig. \ref{fig_kappa_l}) and $\text{d}\kappa_l(\omega)/\text{d}\omega$ (Fig. \ref{fig_spectral_figure}) is obtained when taking into account the dispersive nature of phonons, without changing the type or strength of the dominant scattering mechanisms. In particular, the simple model phonon dispersion as used here in combination with the empirical expressions for the different scattering processes, but without any further fitting parameter, reproduces the cost-intensive DFT results both on a microscopic and macroscopic level. We chose a third-order polynomial to mimic a more realistic phonon dispersion. Analytical functions with more free parameters may describe the actual phonon dispersion curves even better and may change the calculated $\kappa_l$ somewhat. One should, however, bear in mind that also the relaxation times for the different scattering processes given in Table \ref{scattering_processes} are based on simple, empirical expressions, so that further refinement of the phonon dispersion without modification of the phonon scattering may quickly cross the limits of the chosen model.
%While the present discussion focuses on the comparison of different models for $\kappa_l$ at \unit[300]{K}, we emphasize that also the temperature dependence of experimental values of $\kappa_l$ is better reproduced using the more refined models (ii) and (iii), while (i) predicts a rather temperature independent $\kappa_l$ for the parameters used here.

Finally, we note that both the improved models (ii) and (iii) require phonon dispersion curves of the material under study. The experimental determination of these dispersion curves can be challenging, due to required sample dimensions and limited access to scattering facilities, while computational methods nowadays routinely provide reliable data. As compared to first principle calculations of thermal transport properties, the phonon dispersion curves required for model (ii) and (iii) are computationally much less demanding, as only harmonic terms have to be considered \cite{Togo2015a}. 

\section{Conclusion}
We conclude that the Callaway model with implementation of the effects of phonon dispersion is a simple but useful model to analyze the lattice thermal conductivity $\kappa_l$ of a material. In contrast, the Callaway model in the often used simplification of Eq. \eqref{kappa_callaway} is generally too simple to deduce the dominant phonon scattering processes at elevated temperatures, and can result in misleading strategies to optimize $\kappa_l$. {\color{black} We have shown, that combining a simple polynomial phonon dispersion relation and empirical expressions for relevant scattering processes without any adjustable parameters reproduces $\kappa_l$ of a $X$NiSn model system both qualitatively and quantitatively.} This model can easily be extended to understand $\kappa_l$ of a wide range of other material systems, as long as information on the phonon dispersion is available.

\begin{acknowledgement}
This work was funded by the Research Council of Norway (THELMA, 228854). MS is grateful for enlightening discussions with Bin He and Joseph Heremans during his stay at the Ohio State University.
\end{acknowledgement}
\bibliographystyle{unsrt}
%\bibliography{M:/literature/HalfHeusler/Half_Heusler}

\end{document}